\documentclass[%
 reprint,
 superscriptaddress,
 amsmath,amssymb,
 aps,
 prb,
]{revtex4-2}

\usepackage{times}
\usepackage{easybmat}
\usepackage{graphicx}
\usepackage{dcolumn}
\usepackage{bm}
\usepackage{hyperref}
\usepackage{xcolor}
\usepackage{verbatim}

\begin{document}

\preprint{APS/123-QED}

\title{Surface density-of-states on semi-infinite topological photonic and acoustic crystals}

\author{Yi-Xin Sha}
\affiliation{Department of Electronics, Peking University, Beijing 100871, China}
\author{Bo-Yuan Liu}
\affiliation{Institute of Physics, Chinese Academy of Sciences/Beijing National Laboratory for Condensed Matter Physics, Beijing 100190, China}
\author{Hao-Zhe Gao}
\affiliation{Institute of Physics, Chinese Academy of Sciences/Beijing National Laboratory for Condensed Matter Physics, Beijing 100190, China}
\affiliation{School of Physical Sciences, University of Chinese Academy of Sciences, Beijing 100049, China}
\author{Heng-Bin Cheng}
\affiliation{Institute of Physics, Chinese Academy of Sciences/Beijing National Laboratory for Condensed Matter Physics, Beijing 100190, China}
\affiliation{School of Physical Sciences, University of Chinese Academy of Sciences, Beijing 100049, China}
\author{Hai-Li Zhang}
\affiliation{School of Electronic Science and Engineering, University of Electronic Science and Technology of China, Chengdu 611731, China}
\author{\\Ming-Yao Xia}
\email{myxia@pku.edu.cn}
\affiliation{Department of Electronics, Peking University, Beijing 100871, China}
\author{Steven G. Johnson}
\affiliation{Department of Mathematics, Massachusetts Institute of Technology, Cambridge, MA 02139, USA}
\affiliation{Department of Physics, Massachusetts Institute of Technology, Cambridge, MA 02139, USA}
\author{Ling Lu}
\email{linglu@iphy.ac.cn}
\affiliation{Institute of Physics, Chinese Academy of Sciences/Beijing National Laboratory for Condensed Matter Physics, Beijing 100190, China}
\affiliation{Songshan Lake Materials Laboratory, Dongguan, Guangdong 523808, China}

\date{\today}

\begin{abstract}
Iterative Green's function, based on cyclic reduction of block tridiagonal matrices, has been the ideal algorithm, through tight-binding models, to compute the surface density-of-states of semi-infinite topological electronic materials. In this paper, we apply this method to photonic and acoustic crystals, using finite-element discretizations and a generalized eigenvalue formulation, to calculate the local density-of-states on a single surface
of semi-infinite lattices.
The three-dimensional~(3D) examples of gapless helicoidal surface states in Weyl and Dirac crystals are shown and the computational cost, convergence and accuracy are analyzed.
\end{abstract}

\maketitle


Topological classical waves are an exciting focus of recent research~\cite{LL-NPhoton-2014,ZX-CommunPhys-2018,OT-RMP-2019,MG-NRevPhys-2019} whose key features are gapless and robust topological surface states at the interfaces. However, there has been no well-established numerical methods to compute a single topological interface state, between semi-infinite 3D bulk crystals, in these lattices made of continuous material systems. In this work, we adopt the block-tridiagonal iterative Green's function method to achieve this goal.

The current dominant approach for calculating these states has been to compute the frequency eigenvalues (band structures) of finite-thickness supercells~(``slabs''). Although easy to understand and implement, this slab method has several limitations. First, there are \emph{two} surfaces on each slab. One has to disentangle the two surface states by checking their wave functions. Second, large supercells may be required to minimize the coupling between surface states localized on opposite surfaces, greatly increasing the computational costs. Third, it is not convenient to obtain iso-frequency cuts of the band diagram, which is required to verify the topological properties such as surface arcs and to compare with the related field-scan experiments~\cite{YB-Science-2018,HB-PRL-2020}.

It would be ideal to compute the states of a single surface on semi-infinite bulk cells. The effective approach is to compute the local-density-of-states (LDOS) on the surface through the Green's functions~\cite{EEN-Book-2006,NL-Book-2006,OA-Book-2013,CWC-PIER-2019}. Many techniques ~\cite{VJ-JPCM-2004} have been developed for the Green’s functions in semi-infinite systems including the recursive method~\cite{HR-JPCS-1972,HR-SSP-1980}, transfer-matrix method~\cite{LDH1-PRB-1981,LDH2-PRB-1981,LDH3-PRB-1981}
and iterative method~\cite{SMPL-JPFM-1984,SMPL-JPFM-1985}. All methods divide the semi-infinite bulk into layers below the surface.

The recursive method writes the Green's function in the form of a continued fraction by relating the neighbouring layers. Only a relatively small matrix, describing each layer, is inverted in each recursion and the effective system size grows layer by layer. For example, a recursive scheme based on a finite-difference discretization was used to compute the edge mode of a semi-infinite two-dimensional~(2D) photonic crystal~\cite{RAI-PRB-2005}. The recursive method is quite general in that each bulk layer can be distinct, i.e. the medium need not be periodic in the direction orthogonal to the surface.

If the semi-infinite bulk cells are all identical, which is the case for periodic lattices, more efficient methods have been developed such as the transfer-matrix and iterative Green's function methods. The transfer-matrix method~\cite{LDH1-PRB-1981,LDH2-PRB-1981,LDH3-PRB-1981} relates the Green's functions of every two neighboring layers with a transfer matrix. By diagonalizing the matrix and obtaining the eigensolutions, the surface Green's function, surface band structure and surface wave functions can all be constructed. For example, the plane-wave transfer-matrix method was used to study the transmission and edge modes in 2D semi-infinite photonic crystals~\cite{LZY-PRB-2003,CM-JOSAA-2008}.
Unfortunately, the transfer matrix is non-Hermitian and the eigenvector basis can be ill-conditioned near an exceptional point~\cite{PA_OE_2017}. Furthermore, solving eigenvectors is much slower than matrix inversions and the size of the transfer matrix is twice as large as that of a single repeating bulk layer.

\begin{figure*}[th!]
\centering
\includegraphics[width=\linewidth]{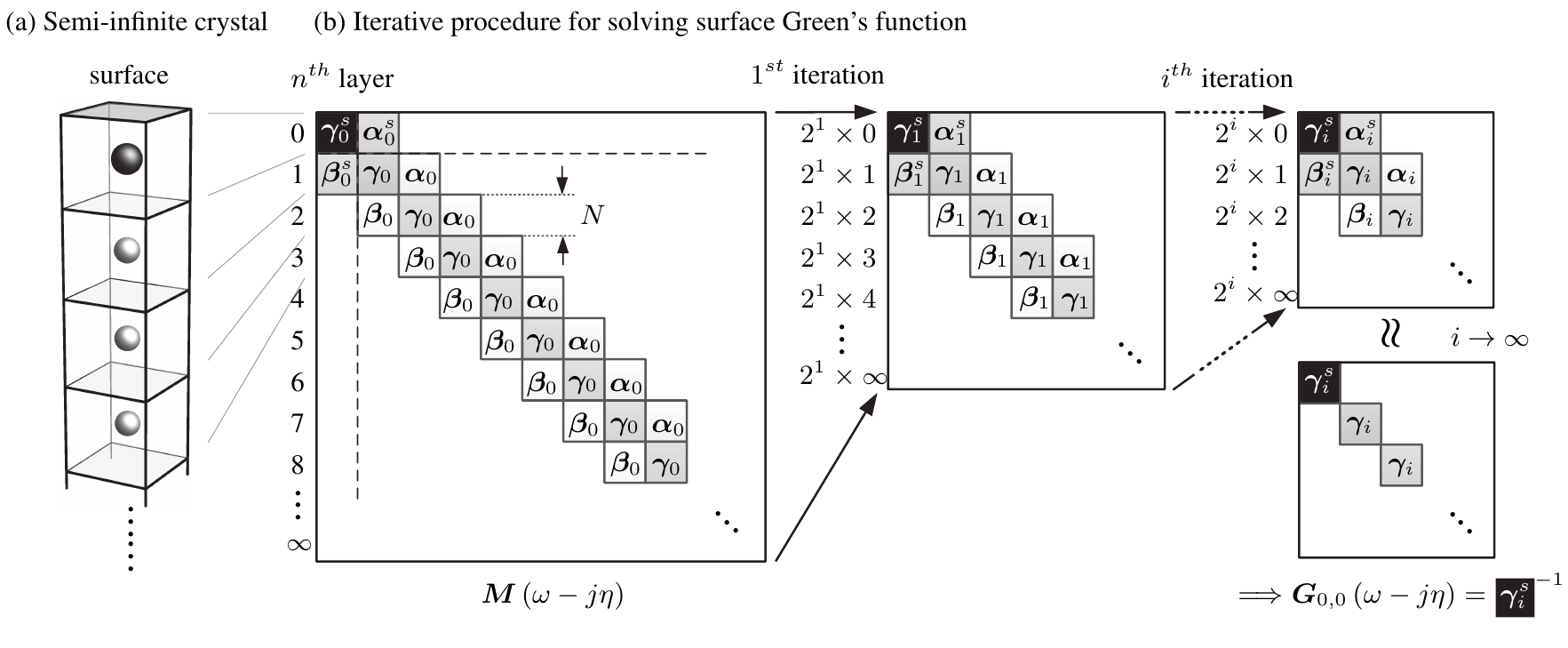}
\caption{\label{fig: schematic}The iterative Green's function method.
(a) Schematic of a semi-infinite crystal. Each cell represents a crystal layer and the top gray surface is the surface boundary. The surface cell can be different from the rest. (b) Iterative procedure for solving the surface Green's function. The block elements~($\bm{\alpha}$, $\bm{\beta}$ and $\bm{\gamma}$) are relabeled from $\bm{M}_{i,j}$ in Eq.~(\ref{eq:eigenmatrix}) and Eq.~(\ref{eq:parameter_0}). After each iteration, half number of the blocks are eliminated and the matrix size halves. When the off-diagonal blocks are sufficiently small, after enough iterations, the Green’s function can be obtained by inverting the remaining diagonal blocks.}
\end{figure*}

The iterative Green's function method~\cite{SMPL-JPFM-1984,SMPL-JPFM-1985} is the most efficient and is the one used in this work. The basic idea illustrated in Fig.~\ref{fig: schematic} is to relate the Green's functions of every even layers, by removing the odd ones, so that the surface layer couples with the $2^i$ layers after $i$ iterations. As a result, the surface Green's function quickly decouples spatially from the bulk and can be solved independently. Historically, similar iterative technique was proposed for rapidly solving linear systems composed of block-cyclic tridiagonal (block-Toeplitz) matrices~\cite{BBL-SJNA-1970,HD-SJNA-1976,ZF-Book-2006,RMG-CSD-2012}. Recently, it has been a standard method for evaluating topological surface states in electronic systems through simplified tight-binding models in a standard eigenvalue problem~\cite{PDE-PhDthesis-2008,ZH-NPhys-2009,WQ-CPC-2018,YC-NPhys-2019}.

We note that the semi-infinite periodic media can also be simulated by implementing the outgoing (radiation) boundary conditions in the bulk medium. Unfortunately, the most popular numerical techniques for absorbing boundaries, the perfectly matched layers~(PML), fails in periodic media~\cite{0A-OE-2008}. The difficulty lies in the fact that the analytical wave solutions in the periodic medium are not known in general. In this regard, methods like nonlinear coordinate transforms~\cite{LP-JOSAA-2005}, adiabatic absorbers~\cite{0A-OE-2008} and Dirichlet-to-Neumann approach~\cite{JP-CPC-2006} have been developed and numerical examples are all given in 2D.

In this paper, we implement the iterative Green's function method for photonic and acoustic crystals through regular finite-element meshing in a generalized eigenvalue problem. Topological surface states of 3D Weyl and Dirac crystals are calculated for demonstration and we discuss the computational performance in the end.

\section{LDOS and Green's Function}
The source-free Maxwell's equations can be written as a frequency-domain eigenproblem for electric field $\bm{E}$. The sound wave equation in fluids can also be expressed in the same form for acoustic pressure $p$. Both governing equations for electromagnetics and acoustics can be unified into a generalized eigenvalue problem with $\omega^2$ as the eigenvalue in Eq.~(\ref{eq:eigenvalue problem}).
\begin{equation}
\left.
\begin{aligned}
\nabla  \times \left( {\frac{1}{{\mu}} \cdot \nabla  \times \bm{E}} \right) = {\omega^2} \varepsilon \cdot \bm{E} \; \\
\nabla  \cdot \left( {\frac{1}{{{\rho}}} \cdot \nabla p} \right) =- {\omega^2}\frac{p}{{{K}}} \;
\end{aligned}
\right\}
\; \bm{A} {\bm{u}} = {\omega^2} \bm{B} {\bm{u}} ,
\label{eq:eigenvalue problem}
\end{equation}
where $\varepsilon$, $\mu$, $\rho$ and $K$ are the permittivity, permeability, mass density and bulk modulus of the material,  respectively. $\bm{A}$ is the differential operator ($\nabla  \times \frac{1}{{\mu}} \nabla  \times$ or $\nabla  \cdot \frac{1}{{{\rho}}} \nabla$),
$\bm{B}$ is the material parameter [$\varepsilon$ or $-\frac{1}{K}$], and ${\bm{u}}$ is the eigenstate ($\bm{E}$ or $p$).

Green's function, the solution to a differential equation excited by a Dirac's delta source, in our system is
\begin{equation}
\left( {\omega^2} \bm{B} - \bm{A} \right) \bm{G} \left( \omega \right)  = \bm{M} \left( \omega \right) \bm{G} \left( \omega \right) = \bm{I} ,
\label{eq:Green's function}
\end{equation}
where $\bm{I}$ is the identity operator.

LDOS, the key physical quantity to compute in this work, describes the response of a point source (the power emitted by a dipole). It can be written in terms of imaginary part of $\bm{G}$ by imposing an infinitesimal imaginary frequency $\eta$~\cite{NL-Book-2006,OA-Book-2013,CWC-PIER-2019}. $\eta$ has to be introduced in lossless Hermitian systems, so that the poles in the Green's function~[$\bm{G}( \omega) =\bm{M} (\omega)^{-1}$] broaden to finite values for numerical evaluation. For vector fields, the Green's function is dyadic, one sums the field components by taking the trace:
\begin{equation}
\begin{aligned}
& {\rm LDOS} \left( {\bm{r} \ ;\omega} \right)=   \\ 
& \frac{2\omega}{\pi }{\rm Tr} \bigg{\{} {\rm Im}\left[\bm{B}\left( \bm{r} \right) \cdot \mathop {\lim }\limits_{\eta  \to {0^ + }} \bm{G}\left( \bm{r}, \bm{r} \ ; \omega  - j\eta \right) \right]  \bigg{\}}.
\label{eq:LDOS}
\end{aligned}
\end{equation}
We emphasize that this definition, as well as this work, applies to any frequency-independent material parameters including lossy and gyrotropic terms. $\bm{B}$ is usually required to be positive definite for LDOS to be a non-negative real number.

\section{Iterative Green's function Method}
By discretizing the system with finite elements~\cite{HB-PhDthesis-2002,JJM-Book-2015,RT-Book-2012}, we obtain the semi-infinite eigenmatrix:
\begin{equation}
\begin{aligned}
& \bm{M}\left( \omega-j\eta \right) =\left( \omega-j\eta \right)^2 \bm{B} - \bm{A}=\\
&\left( \begin{BMAT}{c.cccccc}{c.ccccc}
{{\bm{M}_{0,0}}}&{{\bm{M}_{0,1}}}&{}&{}&{}&{}&{}\\
{{\bm{M}_{1,0}}}&{{\bm{M}_{1,1}}}&{{\bm{M}_{1,2}}}&{}&{}&{}&{}\\
{}&{{\bm{M}_{2,1}}}&{{\bm{M}_{1,1}}}&{{\bm{M}_{1,2}}}&{}&{}&{}\\
{}&{}&{\ddots}&{\ddots}&{\ddots}&{}&{}\\
{}&{}&{}&{{\bm{M}_{2,1}}}&{{\bm{M}_{1,1}}}&{{\bm{M}_{1,2}}}&{}\\
{}&{}&{}&{}&{\ddots}&{\ddots}&{\ddots}\\
\end{BMAT} \right) .
\label{eq:eigenmatrix}
\end{aligned}
\end{equation}
$\bm M$ is block cyclic tridiagonal and the subscripts represent the layer numbers.
The diagonal block ${{\bm{M}_{n,n}}}$ represents the intra-coupling matrix of the $n$-th layer.
The off-diagonal blocks ${{\bm{M}_{n,n+1}}}$ and ${{\bm{M}_{n+1,n}}}$ denote the inter-coupling matrix between the neighbouring layers, whose meshes are joined only at the layer boundaries. These bulk block matrices are identical to ${{\bm{M}_{1,1}}}$, ${{\bm{M}_{1,2}}}$ and ${{\bm{M}_{2,1}}}$, due to the semi-infinite crystal periodicity.
In order to fully accommodate the realistic conditions for the sample surfaces, the surface layer ($n=0$) is assumed here to be \emph{arbitrarily} different from the bulk layers in its thickness, materials and geometry. 

We now derive the surface Green's function $\bm{G}_{0,0}$ using Eq.~(\ref{eq:Green's function}) and (\ref{eq:eigenmatrix}). Multiplying $\bm{M}$ by the zeroth block row of the matrix $\bm{G}$, we get a series of chain equations:
\begin{subequations}
\begin{equation}
\left\{
\begin{aligned}
{-\bm{\gamma} _0^s}{\bm{G}_{0,0}} &=  - \bm{I} + {\bm{\alpha} _0^s}{\bm{G}_{1,0}}\\
{-\bm{\gamma} _0}{\bm{G}_{1,0}} &= {\bm{\beta} _0^s}{\bm{G}_{0,0}} + {\bm{\alpha} _0}{\bm{G}_{2,0}}\\
{-\bm{\gamma} _0}{\bm{G}_{n,0}} &= {\bm{\beta} _0}{\bm{G}_{n - 1,0}} + {\bm{\alpha} _0}{\bm{G}_{n + 1,0}}\;\;  (n\geq2)
\end{aligned}
\right.
\label{eq:chain_0}
\end{equation}
with
\begin{equation}
\begin{aligned}
&{\bm{\alpha} _0} = {\bm{M}_{1,2}} , \ \bm{\alpha} _0^s = {\bm{M}_{0,1}}  ,\\
&{\bm{\beta} _0} = {\bm{M}_{2,1}}  , \ \bm{\beta} _0^s = {\bm{M}_{1,0}}  ,\\
&{\bm{\gamma} _0} = {\bm{M}_{1,1}} , \ \bm{\gamma} _0^s = {\bm{M}_{0,0}}  ,
\end{aligned}
\label{eq:parameter_0}
\end{equation}
\end{subequations}
where ${\bm{\alpha} _0}$, $\bm{\alpha} _0^s$, ${\bm{\beta} _0}$, $\bm{\beta} _0^s$, ${\bm{\gamma} _0}$, $\bm{\gamma} _0^s$ are the changes of notations easy for later iterations and the superscript $s$ denotes the surface layer. By eliminating the odd-layer~(odd-index) Green's functions, we update Eq.~(\ref{eq:chain_0}) as
\begin{subequations}
\begin{equation}
\left\{
\begin{aligned}
&{-\bm{\gamma} _1^s}{\bm{G}_{0,0}} =  - \bm{I} + \bm{\alpha} _1^s{\bm{G}_{2,0}}\\
&{-\bm{\gamma} _1}{\bm{G}_{2,0}} = \bm{\beta} _1^s{\bm{G}_{0,0}} + {\bm{\alpha} _1}{\bm{G}_{4,0}}\\
&{-\bm{\gamma} _1}{\bm{G}_{2n,0}} = {\bm{\beta} _1}{\bm{G}_{2(n - 1),0}} + {\bm{\alpha} _1}{\bm{G}_{2(n + 1),0}}\;\;  (n\geq2)
\end{aligned}
\right.
\label{eq:chain_1}
\end{equation}
with
\begin{equation}
\begin{aligned}
{\bm{\alpha} _{1 }} &= {\bm{\alpha} _{0}}{\left( {{\bm{\gamma} _{0}}} \right)^{ - 1}}{\bm{\alpha} _{0}}  ,\\
\bm{\alpha} _{1 }^s &= \bm{\alpha} _{0}^s{\left( {{\bm{\gamma} _{0}}} \right)^{ - 1}}{\bm{\alpha} _{0}}  , \\
{\bm{\beta} _{1 }} &= {\bm{\beta} _{0}}{\left( {{\bm{\gamma} _{0}}} \right)^{ - 1}}{\bm{\beta} _{0}}  ,\\
\bm{\beta} _{1 }^s &= \bm{\beta} _{0}{\left( {{\bm{\gamma} _{0}}} \right)^{ - 1}}{\bm{\beta} _{0}^s}  , \\
{\bm{\gamma} _{1 }} &= {\bm{\gamma} _{0}} - {\bm{\alpha} _{0}}{\left( {{\bm{\gamma} _{0}}} \right)^{ - 1}}{\bm{\beta} _{0}} - {\bm{\beta} _{0}}{\left( {{\bm{\gamma} _{0}}} \right)^{ - 1}}{\bm{\alpha} _{0}}  , \\
\bm{\gamma} _{1 }^s &= \bm{\gamma} _{0}^s - \bm{\alpha} _{0}^s{\left( {{\bm{\gamma} _{0}}} \right)^{ - 1}}\bm{\beta} _{0}^s .
\end{aligned}
\label{eq:parameter_1}
\end{equation}
\end{subequations}

The Eq.~(\ref{eq:chain_1}) has half number of the equations in Eq.~(\ref{eq:chain_0}) but still remains the same structure as Eq.~(\ref{eq:chain_0}). By repeating this procedure, we obtain the general chain equations:
\begin{subequations}
\begin{equation}
\left\{
\begin{aligned}
&{-\bm{\gamma} _i^s}{\bm{G}_{0,0}} =  - \bm{I} + \bm{\alpha} _i^s{\bm{G}_{{2^i},0}}\\
&{-\bm{\gamma} _i}{\bm{G}_{{2^i},0}} = \bm{\beta} _i^s{\bm{G}_{0,0}} + {\bm{\alpha} _i}{\bm{G}_{{2^{i + 1}},0}}\\
&{-\bm{\gamma} _i}{\bm{G}_{2^i n,0}} = {\bm{\beta} _i}{\bm{G}_{2^i (n - 1),0}} + {\bm{\alpha} _i}{\bm{G}_{2^i (n +1),0}} \;\;  (n\geq2)
\end{aligned}
\right.
\label{eq:chain_n}
\end{equation}
with the iterative relations:
\begin{equation}
\begin{aligned}
{\bm{\alpha} _{i }} &= {\bm{\alpha} _{i-1}}{\left( {{\bm{\gamma} _{i-1}}} \right)^{ - 1}}{\bm{\alpha} _{i-1}} ,\\
\bm{\alpha} _{i }^s &= \bm{\alpha} _{i-1}^s{\left( {{\bm{\gamma} _{i-1}}} \right)^{ - 1}}{\bm{\alpha} _{i-1}} ,\\
{\bm{\beta} _{i }} &= {\bm{\beta} _{i-1}}{\left( {{\bm{\gamma} _{i-1}}} \right)^{ - 1}}{\bm{\beta} _{i-1}} ,\\
\bm{\beta} _{i }^s &= \bm{\beta} _{i-1}{\left( {{\bm{\gamma} _{i-1}}} \right)^{ - 1}}{\bm{\beta} _{i-1}^s} ,\\
{\bm{\gamma} _{i }} &= {\bm{\gamma} _{i-1}} - {\bm{\alpha} _{i-1}}{\left( {{\bm{\gamma} _{i-1}}} \right)^{ - 1}}{\bm{\beta} _{i-1}} \\
&- {\bm{\beta} _{i-1}}{\left( {{\bm{\gamma} _{i-1}}} \right)^{ - 1}}{\bm{\alpha} _{i-1}} ,\\
\bm{\gamma} _{i }^s &= \bm{\gamma} _{i-1}^s - \bm{\alpha} _{i-1}^s{\left( {{\bm{\gamma} _{i-1}}} \right)^{ - 1}}\bm{\beta} _{i-1}^s .
\end{aligned}
\label{eq:parameter_n}
\end{equation}
\end{subequations}
The subscript $i$~($i\geq1$) means the $i$-th iterations, after which the surface Green's function $\bm{G}_{0,0}$ couples with the Green's function $\bm{G}_{2^i,0}$ of the $2^i$ layer. The coupling~(off-diagonal) matrices ${\bm{\alpha} _i^s}$, ${\bm{\alpha} _i}$, ${\bm{\beta} _i^s}$, ${\bm{\beta} _i}$ approaches zero exponentially fast with the iteration number, so that, after a few iterations, the surface Green's function in Eq.~(\ref{eq:chain_n}) equals the inverse of the zeroth diagonal block:
\begin{equation}
{\bm{G}_{0,0}}\left( \omega-j\eta \right) = \mathop {\lim }\limits_{i \to \infty } {\left( {\bm{\gamma} _i^s} \right)^{ - 1}} .
\label{eq:surface Green's function}
\end{equation}
The surface-layer DOS is proportional to the trace of $\bm{G}_{0,0}$. If the surface layer is identical to the bulk ones, which is the case when the iterative Green's function method was first introduced~\cite{SMPL-JPFM-1984,SMPL-JPFM-1985}, the problem is simplified with ${\bm{\alpha} _{i }}=\bm{\alpha} _{i }^s$ and ${\bm{\beta} _{i }}=\bm{\beta} _{i }^s$ in Eq.~(\ref{eq:parameter_n}).

Importantly, this iterative approach, solving surface DOS on \emph{one} semi-infinite crystal, can be extended to solve the surface DOS at the interface between \emph{two} semi-infinite crystals. As a special example, one of them can be semi-infinite air.

\begin{figure*}[!ht]
\centering
\includegraphics[width=\linewidth]{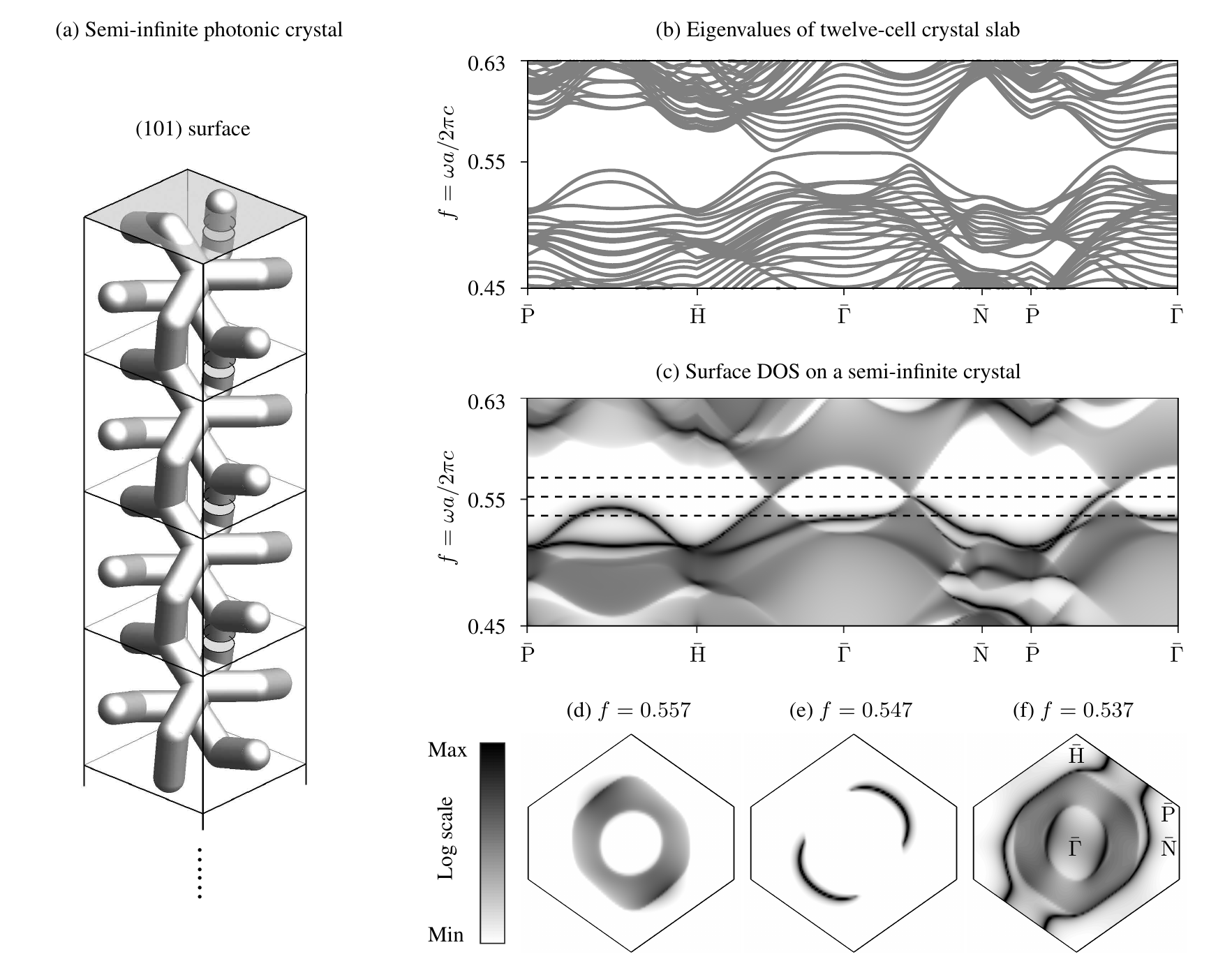}
\caption{\label{fig:photonic example}Surface DOS on a semi-infinite Weyl photonic crystal.
(a) Geometry of semi-infinite dielectric photonic crystals with relative permittivity of 16. The top gray surface is the perfect electric conductor, which satisfies $\hat{\bm{n}} \times \bm{E}=0$ and $\hat{\bm{n}}$ is the surface normal. Here we use cylinders with a radius of 0.10$a$ to approximate the double gyroid structures and an air cylinder with a height of 0.07$a$ to break the inversion symmetry, where $a$ is the lattice constant. (b) Band structure of the twelve-cell photonic crystal slab with perfect electric conductors on both surfaces. The Weyl points are at the normalized frequency ~0.55. (c) Surface DOS on semi-infinite bulk cells. (d), (e) and (f) The iso-frequency cuts at normalized frequencies 0.567, 0.552 and 0.537.}
\end{figure*}

\begin{figure*}[!ht]
\centering
\includegraphics[width=\linewidth]{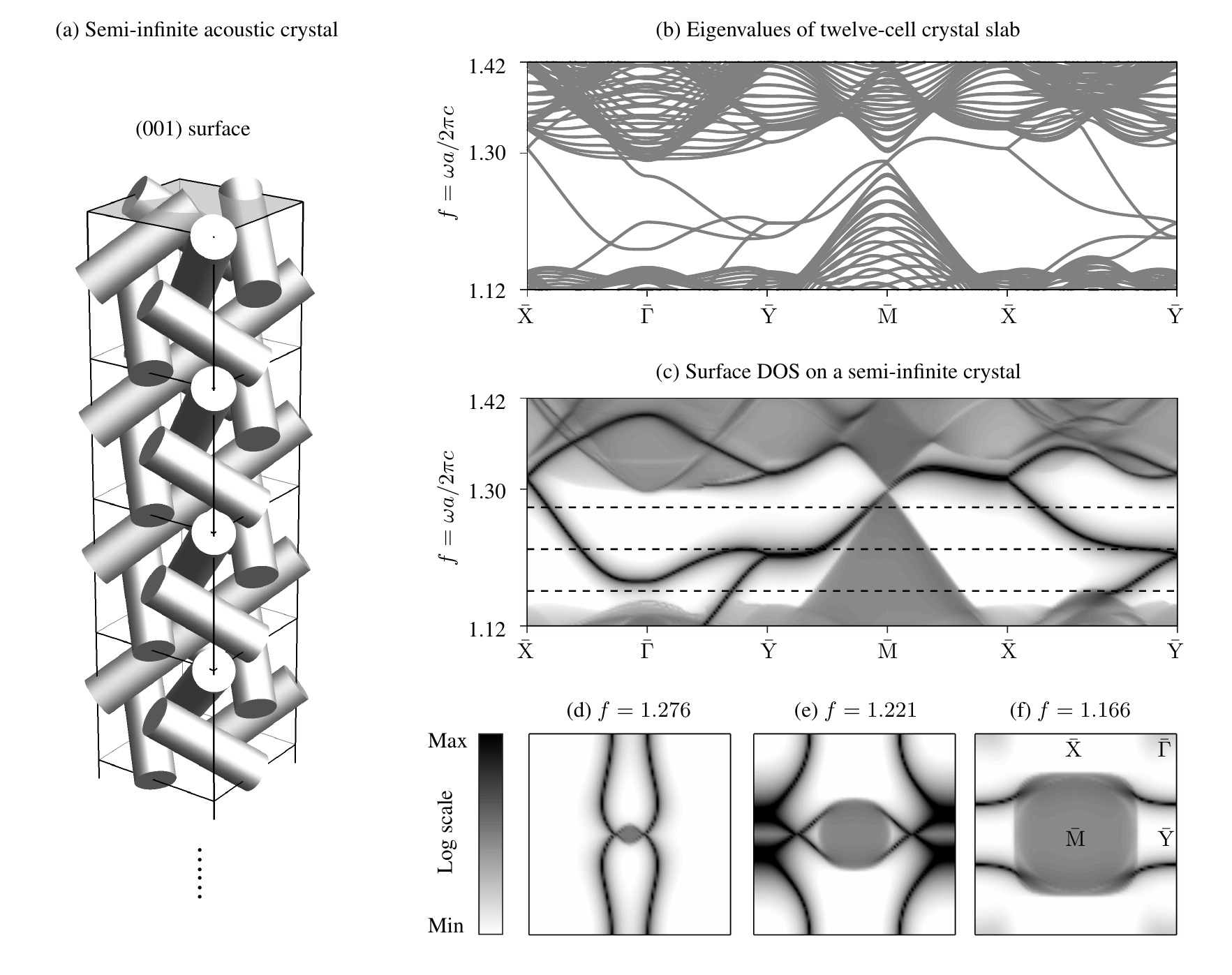}
\caption{\label{fig:acoustic example}
Surface DOS on a semi-infinite Dirac acoustic crystal.
(a) Geometry of the semi-infinite blue-phase-I acoustic crystal. The material is taken as hard wall boundaries in numerics, which satisfies $\hat{\bm{n}} \cdot \nabla p=0$ and $\hat{\bm{n}}$ is the surface normal.
(b) Band structure of twelve-cell acoustic crystal slab with two hard wall boundaries. The Dirac points are at the normalized frequency ~1.30.
(c) Surface DOS on the semi-infinite bulk cells.
(d), (e) and (f) The iso-frequency cuts at normalized frequencies 1.276, 1.221 and 1.166.}
\end{figure*}

\section{Numerical examples}
\label{sec::NumericalExamples}
Using the above iterative Green's function method, we calculate the surface DOS in semi-infinite 3D topological photonic and acoustic crystals with Bloch-periodic boundary conditions, specified by the Bloch wavevector $\bm k$, in the surface-parallel directions. In the surface-normal direction, the crystals are terminated with perfect electrical conductor or hard wall boundary.

The photonic example, in Fig.~\ref{fig:photonic example}, is the double-gyroid dielectric photonic crystal~\cite{LL-NPhoton-2013, LL-Science-2015} having four bulk Weyl points and a single helicoid surface state~\cite{FC-NPhys-2016}. The surface arcs connecting the four projected Weyl points is plotted in Fig.~\ref{fig:photonic example}(e).

The acoustic example, in Fig.~\ref{fig:acoustic example}, is the blue-phase-I acoustic crystal having two Dirac points and four helicoid surface states protected by the glide symmetries~\cite{HB-PRL-2020}. The two Dirac points projects onto the same point in the surface Brillouin zone attached with four surface arcs, as shown in Fig.~\ref{fig:acoustic example}(d).

The data quality from the semi-infinite crystals are superior, in many ways, than the band structures from the twelve-cell slab calculations in  Fig.~\ref{fig:photonic example}(b) and Fig.~\ref{fig:acoustic example}(b). First, the surface states from the second surface do not exist in the semi-infinite data. Second, as the bulk level spacing vanishes in the semi-infinite data, one can identify the bulk continuum and the gapless Weyl and Dirac points. Third, the LDOS intensity, measuring the field localization on the surface, automatically highlights the surface states and compares directly to experiments of near-field scans.

\section{Computing Efficiency}

\begin{figure*}[!ht]
\centering
\includegraphics[scale=1]{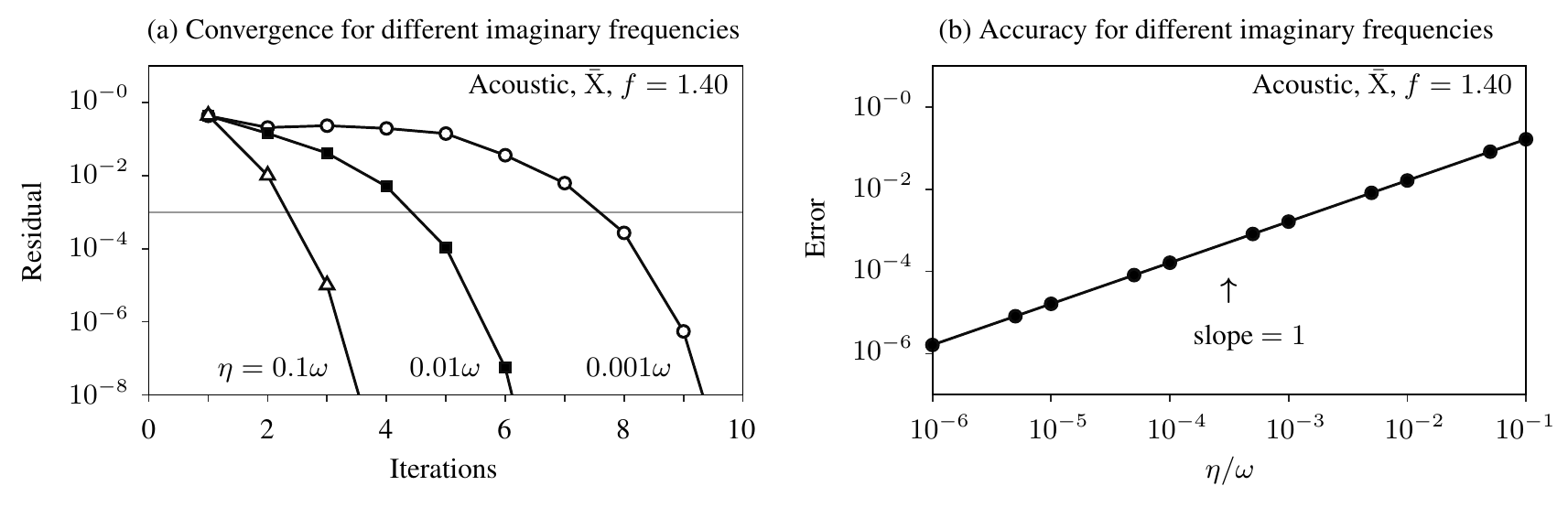}
\caption{\label{fig:convergence}Numerical convergence and accuracy of the acoustic example. (a) Convergence variance with different imaginary frequencies. The residual is defined as $\Vert \bm{\gamma} _{i }^s(\eta)-\bm{\gamma} _{i-1 }^s(\eta)\Vert_F / \Vert\bm{\gamma} _{i-1 }^s(\eta) \Vert_F$, where $i$ is the iteration step and $\Vert \ \Vert_F$ represents the Frobenius norm of a matrix. (b) Accuracy dependence on different imaginary frequencies. The error is defined as $\mathop {\lim }\limits_{i \to \infty } \Vert \bm{\gamma}_i^s(\eta)-\bm{\gamma}_i^s(0^+)\Vert_F / \Vert\bm{\gamma}_i^s(0^+) \Vert_F$.}
\end{figure*}

The total data points are $125 \times 250 $, for the band structures in Fig.~\ref{fig:photonic example}c and  Fig.~\ref{fig:acoustic example}c, and $80 \times 80 $ for the iso-frequency cuts in Fig.~\ref{fig:photonic example}d,e,f and  Fig.~\ref{fig:acoustic example}d,e,f.
The detailed computational costs of each data point $(\bm{k},\omega)$ are listed in Table~\ref{tab:time and memory} for one iteration.
Different numbers of iterations are required to converge for different data points, typically ranging from 3 to 5 iterations when the residual is set to be $10^{-3}$ and $\eta$ is set to be 0.01$\omega$ in Fig.~\ref{fig:convergence}.

Mathematically, introducing the imaginary frequency $\eta$ is equivalent to doing a frequency average of LDOS in a Lorentzian window~\cite{LX_OE_2013}, where $\eta$ determines the broadening and the height of the peaks at the poles.
If we interprete $\eta$ as the quality factor $Q={\omega}/2{\eta}$ of the system, $\eta = 0.01\omega$ means $Q = 50$. We linearly scale the $\eta$ with $\omega$ to ensure the same linewidth broadening across the whole spectrum.
As shown in Fig.~\ref{fig:convergence}, the Green's function converges faster with a larger $\eta$, but  the error also increase linearly with $\eta$.

The bottle neck of this method is the fast growing computational costs with the problem size, which is the number of unknowns of a single cell --- the block submatrix size $N$ as shown in Fig.~\ref{fig: schematic}b. Although the finite-element eigenmatrix $\bm{M}$ is initially sparse, its inverse is not~($\bm{\gamma}^{-1}$ in Eq.~\ref{eq:parameter_1},~\ref{eq:parameter_n},~\ref{eq:surface Green's function}). So faster algorithms designed for sparse matrices cannot help and the memory and time costs grow with ${N^2}$ and ${N^3}$, like a regular matrix problem. 
Using twenty Intel Xeon 2.30 GHz processing cores, it takes 10$\sim$20 days to obtain the results in Fig.~\ref{fig:acoustic example}c and Fig.~\ref{fig:photonic example}c, half$\sim$one day to obtain the results in Fig.~\ref{fig:acoustic example}d,e,f and Fig.~\ref{fig:photonic example}d,e,f. While it takes no more than half a day to obtain the band structures of a supercell slab as shown in Fig.~\ref{fig:acoustic example}b and Fig.~\ref{fig:photonic example}b.

\begin{table}[!ht]
\centering
\caption{\label{tab:time and memory}Time and Memory Costs of the Surface DOS Calculation for a Single Data Point $(\bm{k},\omega)$}
\begin{ruledtabular}
\begin{tabular}{lll}
                                   &  {Photonic}    & {Acoustic}\\ \hline
Problem size $N$                   & 4802           & 3780      \\
Peak memory (GB)                   & 5.2            & 3.4       \\
Time per iteration\footnotemark[1](s)  & 179.1          & 90.5    \\
\end{tabular}
\end{ruledtabular}
\footnotetext[1]{MATLAB run using one 2.30 GHz Intel Xeon Gold 6140 processor core.}
\end{table}

\section{Conclusion}
We implement the iterative Green's function method to calculate the surface density-of-states in semi-infinite photonic and acoustic crystals. The results are highly desirable for studying the topological states in classical systems, despite the drawback of its large computation costs for 3D problems. This method can be further developed to treat frequency-dependent material parameters~\cite{SA_JCP_2005}.

\begin{acknowledgments}
This work was supported by National Natural Science Foundation of China (61531001, 12025409, 11721404, 11974415), by National Key R\&D Program of China (2017YFA0303800, 2016YFA0302400), by the Strategic Priority Research Program (XDB33000000) and the international partnership program (112111KYSB20200024) of the Chinese Academy of Sciences, and by Beijing Natural Science Foundation (Z200008).
\end{acknowledgments}

\nocite{*}

\bibliography{references}

\end{document}